\documentclass[fleqn,usenatbib]{mnras}

\usepackage{newtxtext,newtxmath}

\usepackage[T1]{fontenc}

\DeclareRobustCommand{\VAN}[3]{#2}
\let\VANthebibliography\thebibliography
\def\thebibliography{\DeclareRobustCommand{\VAN}[3]{##3}\VANthebibliography}

\usepackage{graphicx}	
\usepackage{amsmath}	
\usepackage{multirow}

\usepackage{hyperref}
\hypersetup{
    colorlinks=true,
    linkcolor=cyan,
    citecolor=cyan,
    urlcolor=cyan,
    }

\usepackage{orcidlink}

\title[Hen~2-57 and its possible symbiotic core]{Highly bipolar planetary nebula Hen~2-57 and its possible symbiotic core
}

\author[J. Merc et al.]{
J. Merc,$^{1,2}$\thanks{E-mail: jaroslav.merc@mff.cuni.cz}\orcidlink{0000-0001-6355-2468}
J.~Miko\l{}ajewska,$^{3}$\orcidlink{0000-0003-3457-0020}
K.~I\l{}kiewicz,$^{3}$\orcidlink{0000-0002-4005-5095}
A.~Udalski$^{4}$\orcidlink{0000-0001-5207-5619}
\\
$^{1}$Astronomical Institute, Faculty of Mathematics and Physics, Charles University, V Hole\v{s}ovi\v{c}k{\'a}ch 2, 180 00 Prague, Czech Republic\\
$^{2}$Instituto de Astrof\'isica de Canarias, Calle Vía Láctea, s/n, E-38205 La Laguna, Tenerife, Spain\\
$^{3}$Nicolaus Copernicus Astronomical Center, Polish Academy of Sciences, Bartycka 18, 00–716 Warsaw, Poland\\
$^{4}$Astronomical Observatory, University of Warsaw, Al. Ujazdowskie 4, 00-478 Warszawa, Poland
}

\date{Accepted 2026 September 16. Received 2026 August 25; in original form 2026 June 23}

\pubyear{\the\year{}}

\begin{document}
\label{firstpage}
\pagerange{\pageref{firstpage}--\pageref{lastpage}}
\maketitle

\begin{abstract}
We present a multi-wavelength analysis of the highly bipolar object Hen~2-57 based on new optical spectroscopy from the Southern African Large Telescope (SALT), complemented by archival imaging, and long-term time-series photometry from OGLE and NEOWISE. The nebula exhibits a pronounced hourglass morphology with a symmetry axis at a position angle of $110\pm5^{\circ}$ and an average lobe opening angle of $\sim$$53^{\circ}$. We also identify pronounced north--south brightness asymmetry, with the southern parts of lobes consistently brighter. The spectral energy distribution shows a strong mid-infrared excess that cannot be reproduced by a single-temperature blackbody, indicating the presence of a dense circumstellar dust structure, likely in the form of a disk or torus, analogous to that observed in M 2-9. Optical spectroscopy reveals a complex kinematic structure in which the central core is redshifted relative to both the eastern ($55\pm4\ \mathrm{km\,s^{-1}}$) and western ($46\pm3\ \mathrm{km\,s^{-1}}$) lobes. While standard diagnostic diagrams place the extended lobes in the planetary nebula regime, emission-line diagnostics of the central core are consistent with those of symbiotic stars. The same region also exhibits a steep Balmer decrement, indicative of significant internal dust extinction. Long-term OGLE photometry shows no evidence for coherent stellar pulsations, whereas NEOWISE photometry reveals a gradual mid-infrared fading of $\sim$0.25 mag between 2010 and 2024. Together, these properties support a scenario in which Hen~2-57 hosts a dust-enshrouded D-type symbiotic central engine actively shaping the surrounding bipolar nebula, making it a close analogue of M 2-9.
\end{abstract}




\section{Introduction}

Planetary nebulae (PNe) represent a short but crucial phase in the late evolution of low- and intermediate-mass stars ($\sim$1--8\,M$_{\odot}$). As stars leave the asymptotic giant branch (AGB), they expel their outer envelopes through intense stellar winds, which are subsequently ionized by the hot central star on its way to becoming a white dwarf. The resulting ionized nebulae provide key insights into stellar mass loss, chemical enrichment of the interstellar medium, and the final evolutionary pathways of stars \citep[see, e.g.][]{2022PASP..134b2001K,2022FrASS...9.5287P,2024Galax..12...39K}.

Although the textbook view 
once described PNe as predominantly spherical, modern imaging surveys have revealed that a majority of them display 
highly aspherical and often bipolar morphologies. The origin of these asymmetries has been a matter of long-standing 
debate. Increasing evidence points to the crucial role of binarity, where interactions with a close or wide companion 
can shape the outflows \citep[][]{2017NatAs...1E.117J,2019ibfe.book.....B}. Understanding the 
role of binarity in PN formation is therefore central not only for explaining the observed diversity of morphologies, 
but also for constraining the physics of binary interaction and the late stages of stellar evolution. 

Interestingly, morphologically similar bipolar structures are also observed around symbiotic stars, where the mass donor is still present and actively interacting with a compact companion \citep[see reviews on symbiotic stars by][]{2012BaltA..21....5M, 2025Galax..13...49M}. Extended nebulae around symbiotic systems act as direct records of ongoing mass exchange and outflow, offering a complementary perspective on the same physical mechanisms thought to operate in PNe. Although relatively rare among symbiotics, well-studied examples include R Aqr \citep{2018A&A...616L...3B,2024MNRAS.532.2511S,2025A&A...704L..11L}, Hen 2-104 (“Southern Crab”; \citealt{2001ApJ...553..211C}), Hen 2-147 \citep{1999A&A...348..978C}, and BI Cru \citep{1992A&A...265L..37S}. These objects exhibit high-velocity polar outflows, often reaching several hundred km\,s$^{-1}$, and share remarkable morphological and kinematical similarities with bipolar PNe, suggesting a strong evolutionary and physical connection between the two populations \citep{2000ASPC..199..175C}. In some cases, bipolar planetary nebulae such as M2-9 have even been proposed to host symbiotic-like central systems \citep[e.g.,][]{2011A&A...529A..43C,2022PASJ...74..594D}, although the evidence remains ambiguous in many cases (see, e.g., Sab 53 or Sab 61, \citealt[][]{2021MNRAS.508.1599S}; M 1-91, \citealt[][]{2010RMxAA..46..221T}).

A key distinction from normal PNe is that in symbiotic systems, the shaping processes can be observed in real or near-real time. Precessing jets, episodic mass ejections, and equatorially focused winds have been directly detected and are known to evolve on timescales of months to years (e.g., \citealt{1994A&A...287..154P}, \citealt{2025A&A...704L..11L}). This contrasts with PNe, where the nebular structure represents an integrated fossil record of past mass-loss episodes. Symbiotic nebulae, therefore, provide a unique laboratory for constraining the timing, geometry, and physical conditions of outflow shaping while the donor star is still actively transferring mass. Understanding these processes is essential for linking binary interaction physics to the formation of asymmetric nebulae, and ultimately to the final mass of the white dwarf and the chemical enrichment of the interstellar medium.

In this work, we focus on Hen~2-57, which was first classified as a PN in the objective prism survey of \citet{1967ApJS...14..125H}. 
In the same year, \citet{1967ApJS...14..154W} described it as a peculiar object, consisting of two triangular 
structures of dimensions $15 \times 13$ arcsec that meet at a knot or bright central star. 
Further morphological studies confirmed its extended nature. \citet{1968BAICz..19....1K} reported an axis ratio of 
about 0.3 between the minor and major axes, while \citet{1987A&AS...68...51S} also noted its extended appearance. 

Despite its interesting morphology, Hen~2-57 has received relatively little attention in the literature since then. 
The central source was proposed to be a possible symbiotic star by \citet{1994MNRAS.271..257K}, based on the high density of the core. 
The flux measurements reported in that work were later used by \citet{2005A&A...433..579P}, who placed the object 
in the diagnostic [\ion{O}{iii}] diagram, where its position appeared consistent with the PN locus. 
Hen~2-57 was not included among symbiotic candidates in the catalog of \citet{2000A&AS..146..407B}, but it does 
appear in more recent compilations, such as \citet{2019ApJS..240...21A} and in the New Online Database of Symbiotic Variables 
\citep{2019RNAAS...3...28M,2019AN....340..598M,2026ApJS..285...45M}.  

In this work, we present new spectroscopic observations of \mbox{Hen~2-57} obtained with the Southern African Large Telescope (SALT), 
together with imaging from surveys and available photometric data. The structure of the paper is as follows. 
Section~\ref{sec:observations} describes the observational material, while Section~\ref{sec:results} presents our analysis 
of the nebular morphology, the complex velocity field, the possible symbiotic nature of the core, and the potential connection 
with other "bow tie" or butterfly-shaped PNe proposed to host symbiotic central stars. 
Finally, our conclusions are summarized in Section~\ref{sec:conclusions}.

\section{Observational data}\label{sec:observations}
\subsection{SALT spectroscopy}
We obtained spectroscopic observations of Hen~2-57 on December 22, 2013 using the Robert Stobie Spectrograph \citep[RSS;][]{2003SPIE.4841.1463B,2003SPIE.4841.1634K} on the SALT telescope \citep{2006SPIE.6267E..0ZB,2006MNRAS.372..151O} as part of the project searching for new symbiotic stars \citep[see][]{2014MNRAS.440.1410M,2026MNRAS.545f2146M}. The observations were obtained under programme 2013-2-RSA\_POL-001 (PI: Miszalski) using the same RSS configuration as described by \citet[][]{2014MNRAS.440.1410M}. The exposure time was 1800 s, which was preceded by an additional spectrum with a 60 s exposure to measure H$\alpha$ unsaturated. The
slit position was 113 deg and the
slit width was 1.3 arcsec. 

\subsection{Time-series photometric data}
We searched for publicly available time-series photometry of Hen~2-57. No useful data are available from the All-Sky Automated Survey for Supernovae \citep[ASAS-SN;][]{2014ApJ...788...48S,2017PASP..129j4502K} or the Asteroid Terrestrial-impact Last Alert System \citep[ATLAS;][]{2018PASP..130f4505T,2020PASP..132h5002S,2021TNSAN...7....1S} because of the faintness of the source and severe crowding in the field. The object is also located too far south to be covered by the Zwicky Transient Facility survey \citep[ZTF;][]{2019PASP..131a8003M}. Consequently, the only available optical light curve is provided by the Optical Gravitational Lensing Experiment \citep[OGLE\,III and IV;][]{2008AcA....58...69U,2015AcA....65....1U}. In addition, we analysed infrared light curves from the NEOWISE mission in the $W1$ and $W2$ bands \citep{2011ApJ...731...53M,2014ApJ...792...30M}.

\subsection{Spectral energy distribution}
To construct the spectral energy distribution (SED) of the source, we used photometric measurements from \textit{Gaia} Data Release 3 \citep[DR3;][]{2016A&A...595A...1G,2023A&A...674A...1G}, the Two Micron All Sky Survey \citep[2MASS;][]{2006AJ....131.1163S}, the Wide-field Infrared Survey Explorer \citep[WISE;][]{2010AJ....140.1868W}, the Midcourse Space Experiment \citep[MSX;][]{2001AJ....121.2819P}, AKARI \citep{2007PASJ...59S.369M}, 
and the Herschel InfraRed Galactic Plane Survey \citep[Hi-GAL;][]{2017MNRAS.471..100E} catalogue.

\subsection{Imaging}
To investigate the morphology of the nebula at different wavelengths, we analysed imaging data from several optical and infrared surveys. Optical images were obtained from the VST Photometric H$\alpha$ Survey of the Southern Galactic Plane and Bulge \citep[VPHAS+;][]{2014MNRAS.440.2036D} in the Sloan $u$, $g$, H$\alpha$, $r$, and $i$ filters, and from the DECam Plane Survey \citep[DECaPS;][]{2018ApJS..234...39S} in the $g$, $r$, $i$, $z$, and $Y$ bands. Near- and mid-infrared imaging was obtained from 2MASS ($J$, $H$, and $K_{\rm S}$) and WISE ($W1$--$W4$). We also used archival \textit{Herschel} Space Observatory data obtained with the Photodetector Array Camera and Spectrometer \citep[PACS;][]{2010A&A...518L...1P,2010A&A...518L...2P} in bands centred at 70, 100, and 160~$\mu$m.

\begin{figure}
\centering
\includegraphics[width=\columnwidth]{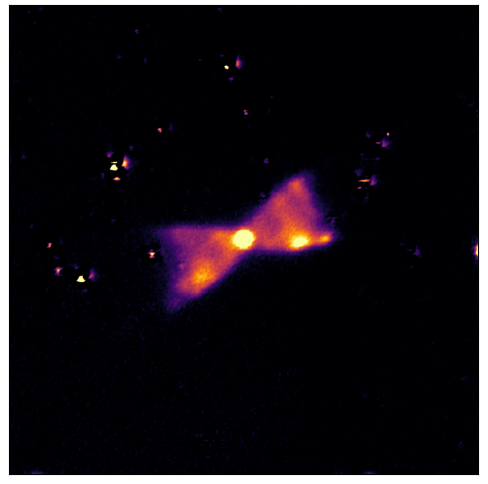}
\caption{VPHAS+ H$\alpha$-r image of the nebula. North is up and east is left. The field shown is 72 $\times$ 72 arcsec.}
\label{fig:halpha-r}
\end{figure}

\begin{figure*}
\centering
\includegraphics[width=\textwidth]{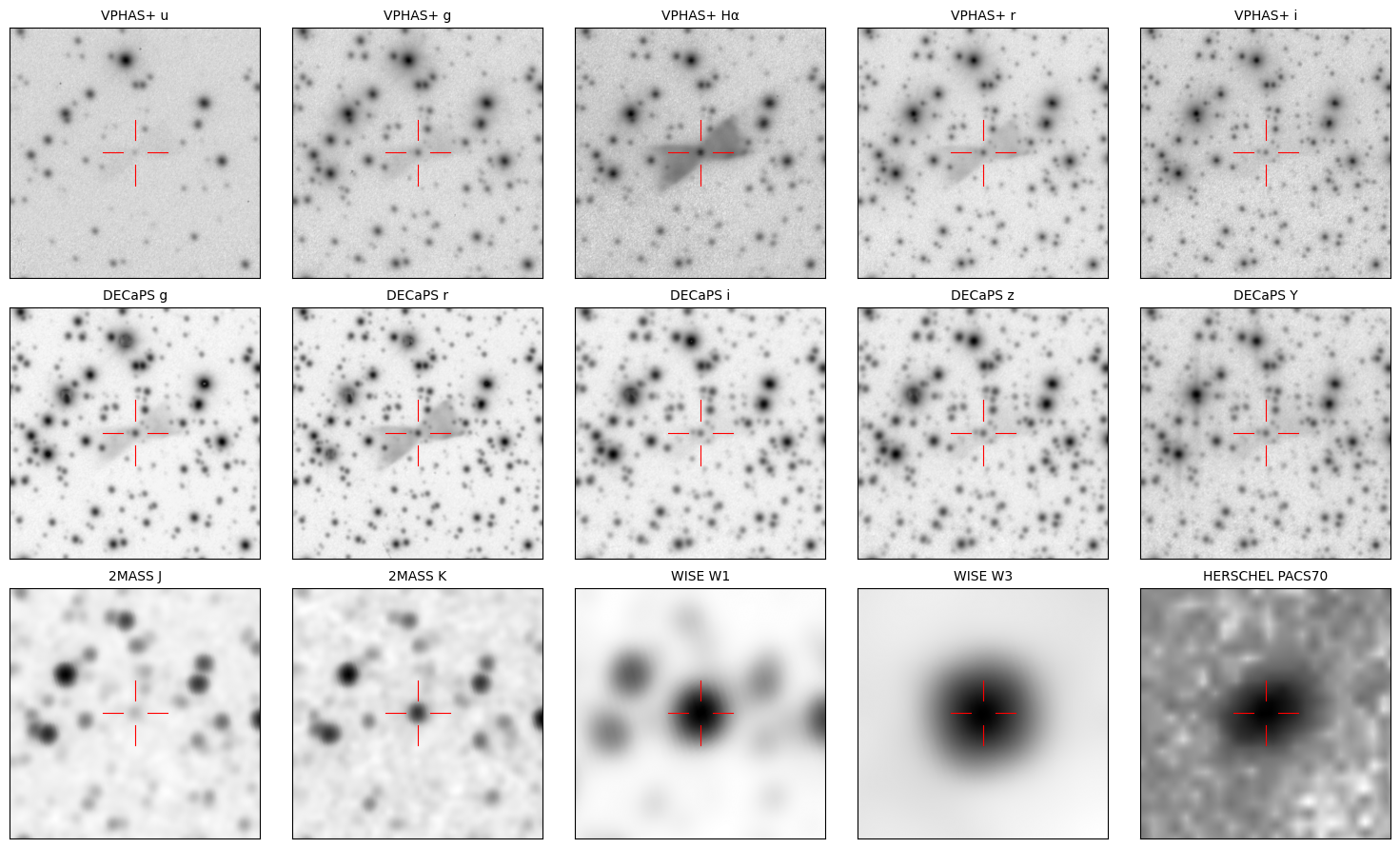}
\caption{VPHAS+, DeCAPS, 2MASS, WISE and HERSCHEL images. In all the images, north is up and east is left. The field shown is 72 $\times$ 72 arcsec.}
\label{fig:mosaic}
\end{figure*}

\section{Results and discussion}\label{sec:results}
\subsection{Optical nebular morphology}\label{sec:optical_morphology}
The morphological resemblance 
between Hen~2-57 and M~2-9 is striking. In Fig. \ref{fig:halpha-r}, we present the continuum-subtracted VPHAS+ H$\alpha$ image of the nebula, where the $r$-band frame has been used to remove the underlying stellar continuum emission (the individual images are shown in Fig. \ref{fig:mosaic}). The resulting image clearly reveals a pronounced bipolar morphology, often described as a bow-tie or hour-glass structure, with a compact central knot coincident with the presumed central source.  

Both lobes are well detected and exhibit a conical geometry with relatively sharp and well-defined outer boundaries, indicative of collimated outflow or strongly anisotropic mass loss. The symmetry axis of the nebula is oriented at a position angle of $110 \pm 5^\circ$. Each bipolar lobe extends to approximately 13\arcsec--14\arcsec from the central source, with transverse widths of about 12\arcsec--14\arcsec. The lobes exhibit opening angles of approximately 55\textdegree{} (eastern lobe) and 51\textdegree{} (western lobe), corresponding to an average opening angle of $\sim$53\textdegree. 

A reliable distance to Hen 2-57 cannot be obtained from the \textit{Gaia} DR3 parallax, which is negative\footnote{Although the RUWE is 1.13, which is within the commonly adopted acceptable range, the astrometric goodness-of-fit statistic of 3.6 indicates a relatively poor fit to the astrometric data.} ($-0.23\pm0.13$ mas; \citealt[][]{2016A&A...595A...1G,2023A&A...674A...1G}), likely due to the faintness of the central source and crowding. We therefore refrain from deriving physical sizes or dynamical ages in this work.

While the global morphology appears largely symmetric, a closer inspection reveals significant small-scale deviations from perfect bilateral symmetry. In particular, the southern portions of both lobes are systematically brighter than their northern counterparts. This brightness asymmetry may reflect local density enhancements, shock interaction with an inhomogeneous interstellar medium, or the presence of compact knots embedded within the outflow. Similar lobe-to-lobe and intra-lobe asymmetries are frequently observed in bipolar nebulae associated with interacting binaries, where jet–wind interaction and precessing outflows produce pronounced brightness gradients and embedded condensations \citep[e.g.][]{1995MNRAS.276..521C, 2015A&A...582A..60C, 2024MNRAS.532.2511S}. In these systems, hourglass morphologies often coexist with collimated fast winds and knotty substructures tracing episodic or directionally variable mass loss from the central binary engine. The observed asymmetry in the present nebula is therefore consistent with the behaviour of bipolar symbiotic nebulae, where departures from axisymmetry are commonly attributed to binary orbital motion, precession of the outflow axis, or localised shock–ionisation effects within the lobes.

The nebular structure is further explored in the far-infrared regime in Sect.~\ref{sec:fir_morphology}, where we find evidence that the same bipolar geometry is also likely traced by cool dust emission. No extended emission is detected in the near- or mid-infrared images (2MASS and WISE; Fig. \ref{fig:mosaic}), indicating that these wavelengths trace only the compact central source rather than the bipolar nebular structure.

\subsection{Far-infrared morphology}\label{sec:fir_morphology}
In contrast to the unresolved appearance of Hen~2-57 in the near- and mid-infrared (2MASS and WISE), the far-infrared emission shows a different spatial behaviour. The Herschel PACS 70~$\mu$m (Fig. \ref{fig:mosaic}) and 160~$\mu$m images reveal extended emission with an apparent size of approximately 36\arcsec$\times$26\arcsec\footnote{No suitable nearby point sources are available to assess the local point-spread function in these images. For reference, the nominal PACS point-spread function has a full width at half maximum of approximately 5.5\arcsec\ at 70~$\mu$m and about 11\arcsec\ at 160~$\mu$m \citep[][]{2024A&A...688A.203M}.}. The emission is elongated along a position angle consistent with that of the optical bipolar nebula (Sect.~\ref{sec:optical_morphology}), indicating a strong morphological connection between the ionized gas and the dust distribution.

The lack of detectable elongation in the near- and mid-infrared suggests that emission at these wavelengths is dominated by compact warm dust in the immediate vicinity of the central source, whereas the far-infrared emission traces a more extended and cooler dust component associated with the bipolar nebular structure. This indicates a pronounced radial and thermal stratification in the dust distribution, with warm dust confined to the inner system and cooler dust occupying the extended lobes or cavity walls.

The spatial coincidence between the FIR morphology and the optical bipolar axis suggests that the large-scale dust reservoir likely follows the same large-scale shape as the ionized nebula, implying that the shaping mechanism may operate on both gas and dust components.

\subsection{Spectral energy distribution}\label{sec:sed}

\begin{figure}
\centering
\includegraphics[width=\columnwidth]{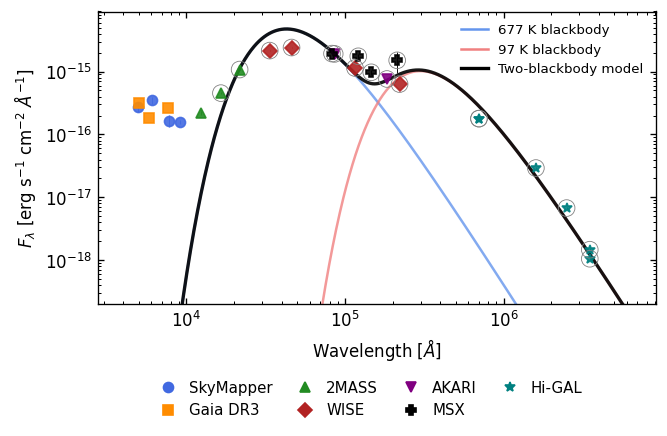}
\caption{{SED of Hen~2-57. Photometric data are from SkyMapper, \textit{Gaia} DR3, 2MASS, WISE, AKARI, MSX, and Herschel. Circled data points indicate those included in the fit with a two-blackbody model. The individual blackbody components are shown in blue and red, while their sum is shown in black.}}
\label{fig:sed}
\end{figure}

It is well visible in Fig. \ref{fig:mosaic} that the central source is a strong infrared emitter. The spectral energy distribution (SED), shown in Fig. \ref{fig:sed}, exhibits a pronounced infrared excess that cannot be reproduced by a single-temperature blackbody, indicating the presence of multiple thermal components associated with circumstellar dust. 

To characterize the observed infrared excess, we fitted the SED with a two-blackbody model. The fit included the \textit{H} and \textit{K} 2MASS data, together with observations from WISE, AKARI, MSX, and Herschel. The optical SkyMapper and \textit{Gaia} DR3 measurements were not included in the fit because they are dominated by the nebular emission (continuum and lines). The best-fitting model consists of a warm dust component with $T_{\rm warm}=677\pm{}$ K and a colder component with $T_{\rm cold}=97\pm{}$ K. These characteristic temperatures are broadly consistent with the range of dust temperatures observed in D-type symbiotic systems, which are commonly associated with substantial amounts of circumstellar dust \citep[e.g., ][]{2010MNRAS.402.2075A}. However, the two-blackbody model provides only a simplified representation of the SED, with $\chi^2=587.8$ for 14 degrees of freedom and a reduced $\chi^2=42.0$. The large value indicates that the observed SED cannot be adequately described by two single-temperature components alone, and more sophisticated dust models incorporating a range of dust temperatures and realistic dust properties are likely required.

In any case, such an SED is also typical for the bipolar nebula M~2-9, and as already mentioned, M~2-9 is widely believed to host 
a symbiotic binary system obscured by a central dust torus \citep[e.g.,][see also \citealt{2011A&A...527A.105L,2014ApJ...780..156W, 2024A&A...692A.151S}]{2011A&A...529A..43C}.

The extremely red near-infrared colours place Hen~2-57 among dust-rich objects occupying the same region of infrared colour--colour diagrams as D-type symbiotic stars and bipolar nebulae with dusty cores. Such colours are difficult to reconcile with a normal evolved planetary nebula central star and instead indicate the presence of a substantial circumstellar dust reservoir.

Across the near- and mid-infrared (2MASS and WISE), the emission is dominated by a compact, unresolved source. No extended emission is detected at these wavelengths, suggesting that they primarily trace warm dust located in the immediate vicinity of the central binary system. In contrast, far-infrared imaging with Herschel reveals spatially extended emission (Sect.~\ref{sec:fir_morphology}), indicating the presence of an additional, cooler dust component on nebular scales.

We note that the photometric data used to construct the SED are not contemporaneous, which is particularly relevant given the long-term mid-infrared variability discussed in Sect.~\ref{sec:variability}. However, this does not affect the main conclusion that the source exhibits a strong infrared excess and requires multiple dust components to reproduce its broadband emission.




\subsection{Radial velocities}

\begin{figure*}
\centering
\includegraphics[width=\textwidth]{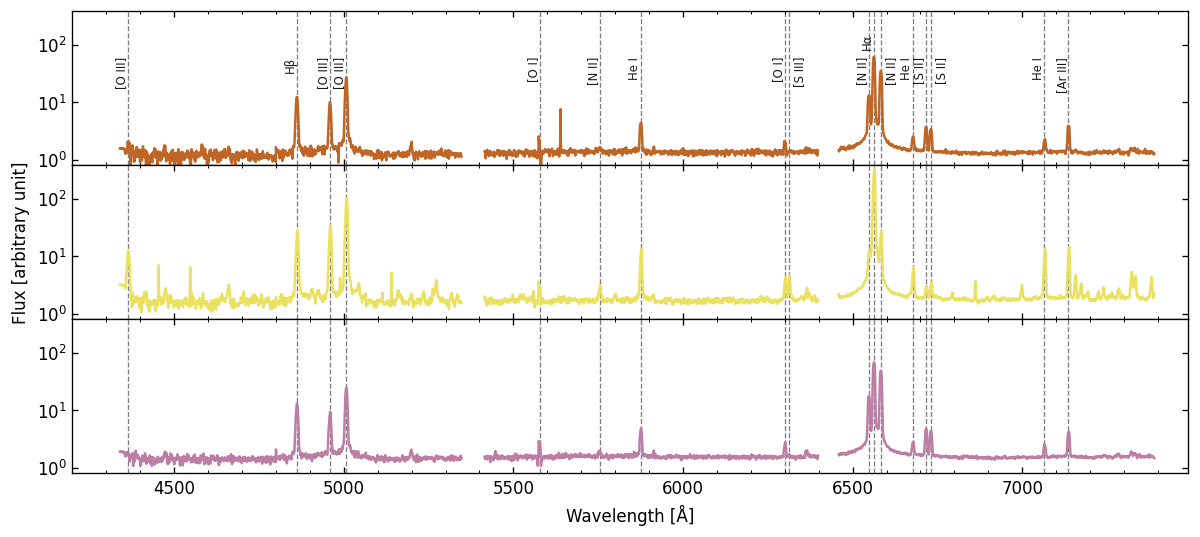}
\caption{SALT spectroscopy of Hen~2-57. Individual panels display the eastern lobe (top), the central core (middle), and the western lobe (bottom). The positions of the main emission lines are indicated. The regions of the original 2D spectra from which data are extracted are shown in Figs. \ref{fig:oIII} and \ref{fig:Ha}.}
\label{fig:spectra}
\end{figure*}

\begin{table*}
\centering
\caption{Emission-line fluxes and heliocentric radial velocities measured in different regions of Hen~2-57. Fluxes are normalized to H$\beta = 100$. }
\label{tab:RV}
\begin{tabular}{lcc cc cc cc cc}
\hline
&
&
&
\multicolumn{2}{c}{Eastern lobe (EL)} &
\multicolumn{2}{c}{Western lobe (WL)} &
\multicolumn{2}{c}{Core (C)} &
\multicolumn{2}{c}{Diff.\ $V_{\mathrm R}$} \\
\cline{4-11}
Emission line &
IP &
$n_{\mathrm{crit}}$ &
Flux &
$V_{\mathrm R}$ &
Flux &
$V_{\mathrm R}$ &
Flux &
$V_{\mathrm R}$ &
C--EL &
C--WL \\
&
[eV] &
[cm$^{-3}$] &
[H$\beta = 100$] &
[km\,s$^{-1}$] &
[H$\beta = 100$] &
[km\,s$^{-1}$] &
[H$\beta = 100$] &
[km\,s$^{-1}$] &
[km\,s$^{-1}$] &
[km\,s$^{-1}$] \\
\hline
$[$O\,{\sc iii}$]$ 4363.21 & 35.1 & $3.3\times10^{7}$ & - & - & - & - & 49 & 119: & - & - \\
H$\beta$ 4861.33           & 13.6 & --                  & 100 & 34 & 100 & 52 & 100 & 120 & 86 & 68 \\
$[$O\,{\sc iii}$]$ 4958.92 & 35.1 & $7\times10^{5}$     & 74 & 28 & 63 & 54 & 119 & 101 & 73 & 47 \\
$[$O\,{\sc iii}$]$ 5006.85 & 35.1 & $7\times10^{5}$     & 300 & 29 & 201 & 42 & 378 & 99 & 70 & 57 \\
$[$N\,{\sc ii}$]$ 5754.59  & 14.5 & $3.2\times10^{7}$   & - & - & - & - & 5 & 75 & - & - \\
He\,{\sc i} 5875.64        & 24.6 & --                  & 25 & 12 & 24 & 24 & 35 & 69 & 57 & 45 \\
$[$N\,{\sc ii}$]$ 6548.06  & 14.5 & $6.6\times10^{4}$   & 106 & 12 & 128 & 5 & 48 & 76 & 64 & 53 \\
H$\alpha$ 6562.82          & 13.6 & --                  & 519 & 4 & 556 & 17 & 1112 & 60 & 56 & 43 \\
$[$N\,{\sc ii}$]$ 6583.29  & 14.5 & $6.6\times10^{4}$   & 293 & 8 & 378 & 5 & 86 & 40 & 32 & 35 \\
He\,{\sc i} 6678.15        & 24.6 & --                  & 9 & 17 & 9 & 20 & 14 & 54 & 37 & 34 \\
$[$S\,{\sc ii}$]$ 6716.40  & 10.4 & $1.5\times10^{3}$   & 19 & 14 & 26 & 15 & 4 & 61 & 47 & 46 \\
$[$S\,{\sc ii}$]$ 6730.80  & 10.4 & $3.9\times10^{3}$   & 16 & 10 & 22 & 17 & 6 & 55 & 45 & 38 \\
He\,{\sc i} 7065.19        & 24.6 & --                  & 7 & 52 & 8 & 48 & 37 & 89 & 37 & 41 \\
$[$Ar\,{\sc iii}$]$ 7135.8 & 27.6 & $4.8\times10^{6}$   & 23 & 43 & 23 & 53 & 39 & 94 & 51 & 41 \\
\hline
$\langle V_{\mathrm R}\rangle_{\rm all}$ &
-- &
-- &
-- & $22\pm4$ &
-- & $29\pm5$ &
-- & $79\pm7$ &
$55\pm4$ &
$46\pm3$ \\
$\langle V_{\mathrm R}\rangle_{\rm low}$ &
-- &
$<10^{5}$ &
-- & $11\pm2$ &
-- & $10\pm3$ &
-- & $58\pm7$ &
$47\pm7$ &
$43\pm4$ \\
$\langle V_{\mathrm R}\rangle_{\rm high}$ &
-- &
$>10^{5}$ &
-- & $33\pm5$ &
-- & $50\pm4$ &
-- & $98\pm7$ &
$65\pm7$ &
$48\pm5$ \\
\hline
\end{tabular}
\end{table*}

The central region of Hen~2-57 exhibits a pronounced redshift relative to both lobes of the nebula: by 55$\pm$4 km\,s$^{-1}$ compared to the eastern lobe and 46$\pm$3 km\,s$^{-1}$ compared to the western lobe. This velocity offset is apparent in all examined emission lines, though it appears more prominent in those associated with high-density regions. No significant radial velocity (RV) difference is detected between the two lobes in low-density lines, whereas a small but measurable offset may be present in lines 
tracing denser material. To minimize possible systematic effects from instrumental wavelength calibration, we primarily rely on differential RV (as listed in the last two columns of Table~\ref{tab:RV}), which provide a more robust comparison between spatial components of the nebula.

In a classical symmetrically expanding bipolar nebula, the two lobes are expected to show roughly equal and opposite radial velocity shifts relative to the central engine. However, in Hen 2-57, the ambient low-density gas displays nearly identical radial velocities in both the eastern ($11\pm2\text{ km s}^{-1}$) and western ($10\pm3\text{ km s}^{-1}$) lobes, suggesting that the major axis of the hourglass is oriented close to the plane of the sky ($i \approx 90^\circ$). Consequently, the apparent net redshift of the central core relative to both lobes is highly unlikely to be a geometric projection effect of the bulk nebular expansion. Instead, this velocity offset implies that the emission profiles of the dense, dusty core are artificially skewed or redshifted. This phenomenon can be explained by severe internal dust stratification and localized extinction along our line of sight—supported by the steep Balmer decrement of the core (see Sect. \ref{sec:fluxes}), which selectively obscures the blueshifted, forward-facing components of the inner core outflows. Alternatively, it could trace an unresolved, asymmetric P-Cygni stellar wind profile from the central engine. A very similar kinematic complexity is seen in M 2-9 \citep[][]{2000A&A...354..674S}, where long-slit spectroscopy revealed widespread redshifted components in the outer loops and dust-scattered profiles. Higher-resolution spatially resolved spectroscopy or integral-field observations would be highly desirable to disentangle the full three-dimensional velocity field of Hen 2-57.

\subsection{Emission line fluxes}\label{sec:fluxes}

\begin{figure*}
\centering
\includegraphics[width=1.5\columnwidth]{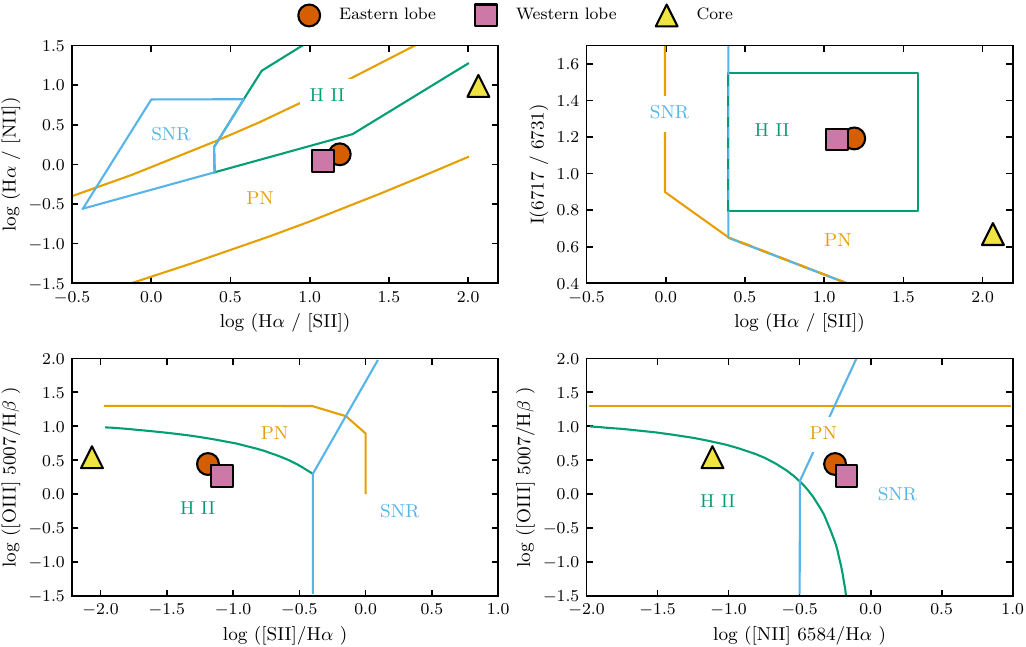}
\caption{Diagnostic diagrams used to distinguish between H\,{\sc ii} regions, PNe, and 
SNRs \citep[][]{2013MNRAS.431..279S}.}
\label{fig:diagrams1}
\end{figure*}

\begin{figure*}
\centering
\includegraphics[width=1.5\columnwidth]{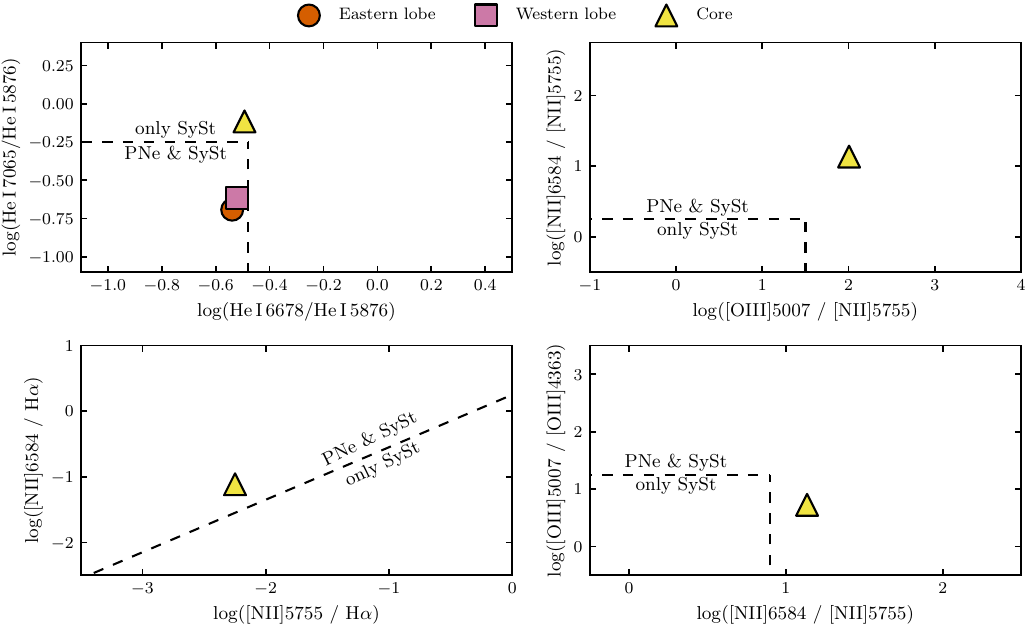}
\caption{Diagnostic diagrams to distinguish PNe and symbiotic stars of \citet[][]{2017A&A...606A.110I}.}
\label{fig:diagrams2}
\end{figure*}

The spectra of Hen~2-57 reveal numerous forbidden emission lines with a wide range of critical densities,
detected in both lobes and in the central region (Fig. \ref{fig:spectra}; Table \ref{tab:RV}). This diversity of excitation conditions indicates a complex
nebular morphology, consistent with the kinematically and morphologically structured nature inferred from the
velocity analysis.

The detection of strong auroral lines [\ion{O}{iii}] $\lambda$4363 and [\ion{N}{ii}] $\lambda$5755 exclusively in the central region supports the presence of very dense gas near the central source.
{We examined the corresponding auroral-to-nebular line ratios following the standard diagnostics of \citet{2006agna.book.....O}. For the core, we measure
$R_{[\ion{O}{iii}]} = (I_{4959} + I_{5007})/I_{4363} = 10.14$
and
$R_{[\ion{N}{ii}]} = (I_{6548} + I_{6584})/I_{5755} = 26.80$.
However, these ratios do not yield a mutually consistent solution for $T_e$ and $n_e$ when interpreted with the standard diagnostic equations. In particular, [\ion{O}{iii}] and [\ion{N}{ii}] trace different ionization zones because of their substantially different ionization potentials, and their respective critical densities also differ significantly. Consequently, the two ratios cannot be straightforwardly interpreted as diagnostics of a single homogeneous gas component. The [\ion{O}{iii}] ratio in particular demonstrates the importance of collisional de-excitation, as its value is far below the low-density prediction for conventional nebular temperatures, which would otherwise require an unrealistically high temperature of $T_e \simeq 1.3\times10^5$~K. At the same time, the [\ion{N}{ii}] ratio is substantially less reliable because both nebular lines are faint, $\lambda$5755 is among the weakest detected lines, and $\lambda$6548 lies on the strong H$\alpha$ wing. Moreover, the measured $I_{6584}/I_{6548}\simeq1.8$ differs significantly from the expected ratio of approximately 3, further illustrating the uncertainty of the [\ion{N}{ii}] flux measurements. We therefore refrain from assigning a formal $T_e$ and $n_e$ to the core from these ratios alone. Instead, the simultaneous presence of the auroral lines and the strong suppression of the nebular [\ion{O}{iii}] and [\ion{N}{ii}] lines provide qualitative evidence for a dense central ionized region. The core $[\ion{S}{ii}]$ doublet ratio ($I_{6716}/I_{6730} = 0.67$) is likewise close to its high-density limit, independently supporting the presence of dense gas.}

{In contrast, neither [\ion{O}{iii}]~$\lambda$4363 nor [\ion{N}{ii}]~$\lambda$5755 is detected in the eastern or western lobes. The $[\ion{S}{ii}]$ doublet ratio ($I_{6716}/I_{6730} \approx 1.18$--$1.19$) indicates low electron densities in the lobes, $n_e \approx 280$--$290\,\text{cm}^{-3}$. Thus, Hen~2-57 exhibits a strong density contrast between the dense central region and the extended lobes. Such ionization and density stratification is also observed in well-studied D-type symbiotic systems such as HM~Sge and V1016~Cyg \citep{1990MNRAS.246...84S}. The observed structure is consistent with a compact, dense ionized region embedded within a much more tenuous extended nebula and provides an important diagnostic of the nature of Hen~2-57. A more detailed determination of the physical conditions and their
spatial variation will
be possible once higher-quality data are available.}

\begin{figure}
\centering
\includegraphics[width=0.8\columnwidth]{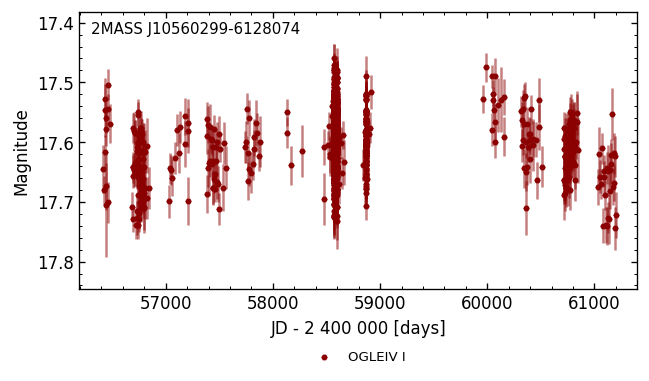}
\includegraphics[width=0.8\columnwidth]{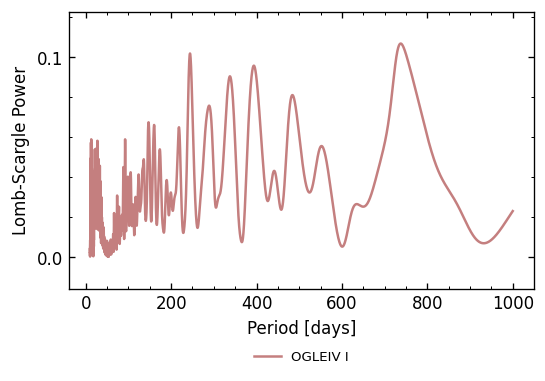}
\caption{Light curve of Hen~2-57 from OGLE survey (upper panel) and the corresponding periodogram (lower panel).}
\label{fig:photometry_ogle}
\end{figure}

\begin{figure}
\centering
\includegraphics[width=\columnwidth]{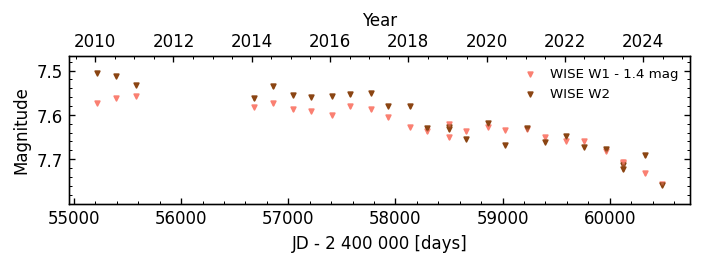}
\caption{AllWISE and NEOWISE \textit{W1} and \textit{W2}-band light curves of Hen~2-57. Average values per observing epoch ($\sim$6-month cadence) are shown.}
\label{fig:photometry_wise}
\end{figure}

Diagnostic diagrams commonly used to distinguish between H,{\sc ii} regions, planetary nebulae (PNe), and
supernova remnants (SNRs) place both the lobes and the central region of Hen~2-57 within the parameter space
typically occupied by PNe (see Fig. \ref{fig:diagrams1}). However, the central region lies somewhat displaced from the positions of the two
lobes, supporting physical differences in excitation or density conditions. These traditional diagnostic plots
do not, however, allow a clear separation between PNe and symbiotic systems. More discriminating diagrams,
involving [\ion{N}{ii}], [\ion{O}{iii}], and He,{\sc i} emission-line ratios
\citep[see Figs.~3 and 4 in][]{2017A&A...606A.110I}, provide further insight.
In the He,{\sc i}-based diagram, the lobes occupy the region typical of PNe, whereas the central region falls
within the locus of known symbiotic stars (Fig. \ref{fig:diagrams2}).

The line fluxes have been corrected for interstellar reddening, adopting $E(B-V) \simeq 0.6$ derived from the observed H$\alpha$/H$\beta$ ratios in the two lobes. The central region exhibits a much steeper Balmer decrement, which may indicate either a higher local density, additional reddening due to circumstellar dust, or a combination of both effects. The latter possibility is consistent with the near-infrared colors of the central source (2MASS J10560299-6128074; $J-H$ = 1.87 mag; $H-K_{\rm s}$ = 2.01 mag), suggesting the presence of a dusty environment. In any case, for consistency when constructing diagnostic diagrams, the same interstellar reddening $E(B-V) \simeq 0.6$ was applied to the core. In this case, if the extra extinction is not fully accounted for, the dereddened line ratios for the core would be underestimated. This would tend to move the core position slightly towards lower excitation in diagrams in Figs. \ref{fig:diagrams1} and \ref{fig:diagrams2}. Even with this conservative approach, the core still falls firmly in the symbiotic locus in the \ion{He}{I} diagnostic diagram, as noted above.

\subsection{Central source and variability}\label{sec:variability}

No strong continuum emission is detected in any of the three regions (Fig.~\ref{fig:spectra}). In particular, there is no evidence of a red continuum that would indicate the presence of a late-type giant, consistent with the characteristics of a typical D-type symbiotics. The absence of clear pulsations in the OGLE light curve (Fig.~\ref{fig:photometry_ogle}) is also not unexpected for D-type symbiotic systems, where Mira pulsations can be heavily obscured by circumstellar dust and may only become apparent at near-infrared wavelengths, as observed in well-studied systems such as RX~Pup and other D-type symbiotics \citep[][]{1999MNRAS.305..190M,2000ASPC..199..431M,2009AcA....59..169G}.

On the other hand, mid-infrared photometry from the WISE satellite (Fig.~\ref{fig:photometry_wise}) reveals a gradual decline in brightness between 2010 and 2024 of approximately 0.25 mag, rather than any coherent periodic variability. Variability at these wavelengths is commonly observed in D-type symbiotic systems and is often attributed to changes in circumstellar dust obscuration. In particular, the W1 band probes wavelengths comparable to the classical $L$ band used in near-infrared studies of D-type symbiotics. Long-term monitoring of systems such as RX~Pup has shown that infrared variability can be dominated by changes in the dust environment, while little or no corresponding variability is present at optical wavelengths \citep[][]{1999MNRAS.305..190M}. The gradual fading observed in Hen~2-57 may therefore reflect changes in the amount or distribution of circumstellar dust. If the dust is concentrated towards the orbital plane, as suggested by the bipolar morphology of the nebula, it could also explain why the presumed Mira component remains undetected in the available optical data.

\subsection{Is Hen~2-57 a planetary nebula with a symbiotic core?}
The classification of Hen 2-57 has remained uncertain for several decades. While it has traditionally been catalogued as a planetary nebula, the possibility that the central source is symbiotic was already suggested by \citet[][]{1994MNRAS.271..257K} based on the unusually dense core. Our new observations provide several independent indications that support this interpretation.

The most compelling evidence comes from the properties of the central region. Unlike the extended lobes, the core occupies the symbiotic star locus in the \ion{He}{i} diagnostic diagram, exhibits a much steeper Balmer decrement, and is the only region where the auroral lines [\ion{O}{iii}] $\lambda$4363 and [\ion{N}{ii}] $\lambda$5755 are detected. Together, these properties indicate a dense and heavily obscured circumstellar environment that differs substantially from the physical conditions in the bipolar lobes.

The infrared properties provide a second line of evidence. The strong infrared excess, extremely red near-infrared colours, and long-term fading detected in the WISE light curves all point to a significant dust component associated with the central source. Such behaviour is characteristic of D-type symbiotic stars, in which the cool giant is embedded in a dense dusty envelope. The absence of detectable optical pulsations does not rule out such a scenario because Mira variability may be hidden by circumstellar obscuration.

The morphology of Hen 2-57 also deserves attention. Its striking resemblance to M 2-9 extends beyond the overall hourglass appearance. Both objects possess compact dusty cores, strong infrared excesses, and evidence for complex internal kinematics. M 2-9 is widely regarded as harbouring a binary central engine, and Hen 2-57 appears to share many of the same observational characteristics.

Although none of the available diagnostics alone provides definitive proof of a symbiotic binary, their combination consistently favours this interpretation. We therefore conclude that Hen 2-57 is most likely a bipolar nebula powered by a dust-enshrouded symbiotic system, making it a strong candidate for another member of the small group of bipolar nebulae occupying the boundary between classical planetary nebulae and symbiotic outflows.

\section{Conclusions}\label{sec:conclusions}
The case of Hen 2-57 highlights the long-standing difficulty in distinguishing PNe from symbiotic stars, as both classes can exhibit strong emission-line spectra and complex morphologies shaped by binary interaction. In symbiotic systems, nebular emission is powered by accretion onto a hot compact star embedded in the wind of a cool giant, whereas in PNe it arises from ionized ejecta of a post-AGB envelope. In practice, these formation channels can produce highly similar observational signatures, making misclassification common when classification relies on morphology or low-resolution optical spectroscopy alone. Hen 2-57 therefore represents a particularly instructive case at the PN--symbiotic boundary.

In this work, we have combined new SALT spectroscopy with archival imaging and long-term optical and infrared photometry to characterise this system. The nebula shows a pronounced bipolar hourglass morphology with an opening angle of $\sim$53\textdegree, together with clear north--south asymmetries in lobe brightness. Its spectral energy distribution exhibits a strong infrared excess, indicative of a substantial circumstellar dust component. Herschel imaging reveals that the bipolar structure of the nebula is also likely present in the far-infrared dust emission, while the near- and mid-infrared emission remains unresolved, indicating a strong wavelength-dependent stratification of the circumstellar environment.

The central region is kinematically and spectroscopically distinct from the extended lobes. It shows high-density emission-line diagnostics, a steep Balmer decrement, and \ion{He}{i} line ratios consistent with symbiotic excitation conditions, while the outer lobes retain properties typical of photoionized planetary nebula gas. In addition, long-term WISE monitoring reveals a gradual decline in mid-infrared flux over the past decade, suggestive of evolving circumstellar dust obscuration.

Taken together, the combined morphological, spectroscopic, kinematic, and photometric evidence favours an interpretation in which Hen 2-57 contains a dust-enshrouded D-type symbiotic nucleus that is actively shaping the surrounding bipolar nebula. Among known objects, it appears to be a close analogue of M 2-9, strengthening its role as a key transitional system in the PN--symbiotic star classification space.


\section*{Acknowledgements}
We thank the referee for comments and suggestions, which improved this paper.
 The research of JaM was supported by the Czech Science Foundation (GACR) project no. 24-10608O. JMik was supported by the Polish National Science Centre (NCN) grant
2023/48/Q/ST9/00138. KI was supported by the Polish NCN grant 2024/55/D/ST9/01713. The OGLE project has received funding from the Polish National Science
Centre grant OPUS-28 2024/55/B/ST9/00447 awarded to AU. The paper is based on spectroscopic observations made with the Southern
African Large Telescope (SALT). Polish participation in SALT is funded
by MNiSW grant No. 2026/WK/02.

.

\section*{Data Availability}
Photometric data discussed in the text are available from the websites of the surveys. Optical spectra are available in the SALT archive.



\bibliographystyle{mnras}
\bibliography{mnras_template} 




\appendix


\begin{figure}
\section{SALT 2D spectra}
\centering
\includegraphics[width=\columnwidth]{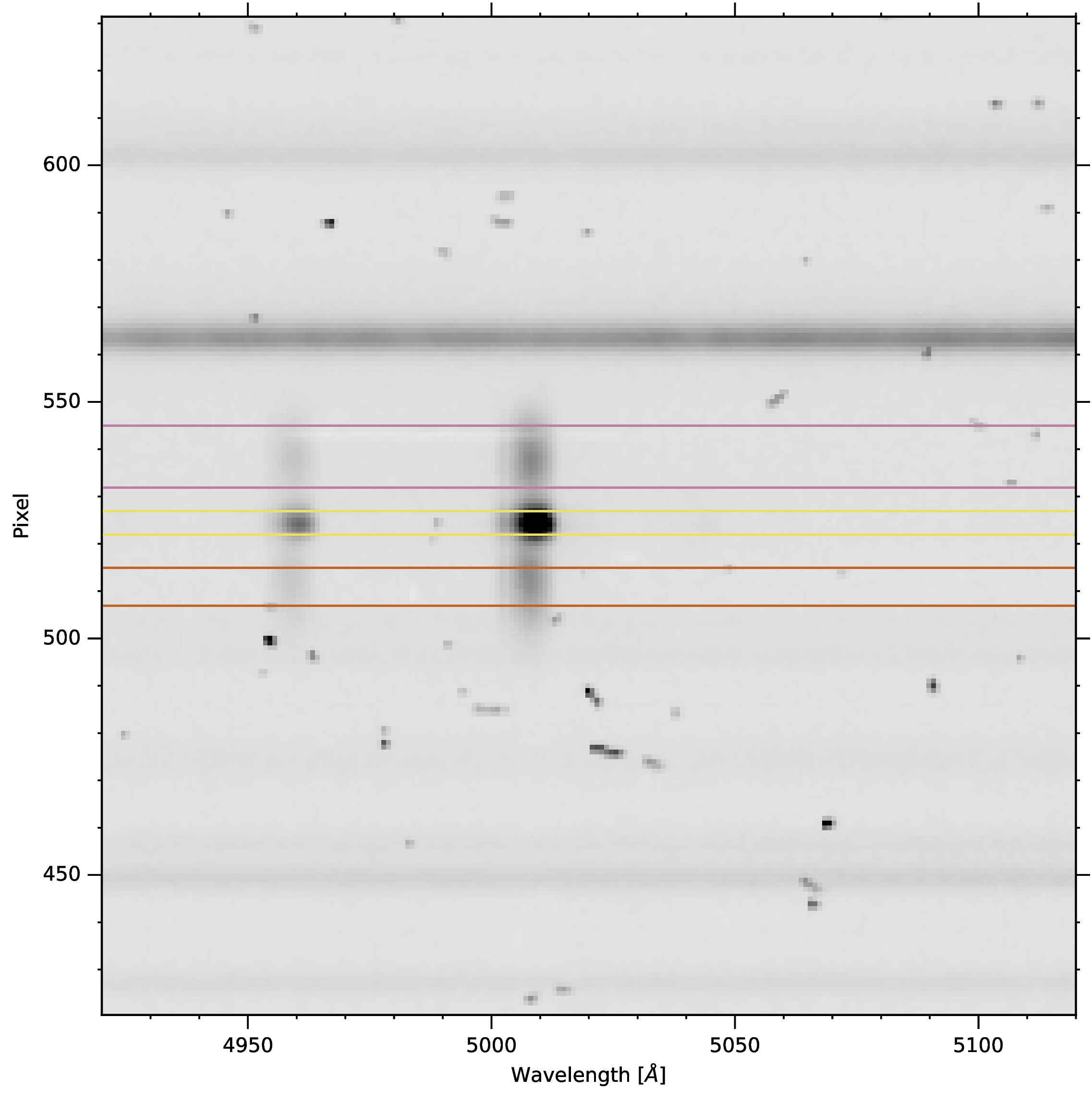}

\caption{SALT 2D spectrum around the [O III] region. The spatial scale along the vertical axis is $0.5068\,\mathrm{arcsec\,pixel^{-1}}$.
The colors are the same as in Fig. \ref{fig:spectra}.}
\label{fig:oIII}
\end{figure}

\begin{figure}
\centering
\includegraphics[width=\columnwidth]{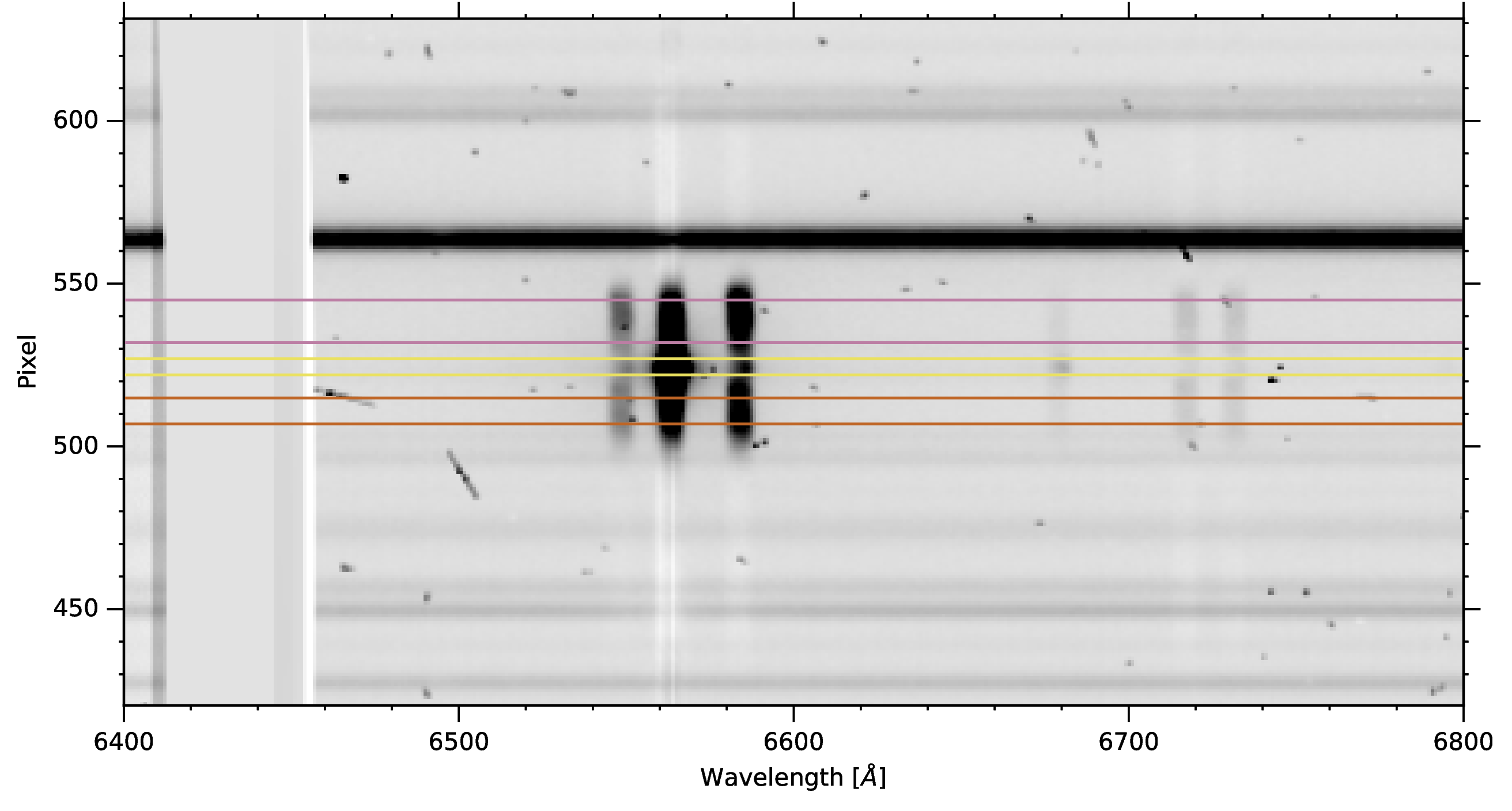}

\caption{SALT 2D spectrum around the H$\alpha$ line region.}
\label{fig:Ha}
\end{figure}





\bsp	
\label{lastpage}
\end{document}